# Integrated cladding-pumped multicore few-mode erbium-doped fibre amplifier for space-division multiplexed communications

H. Chen[1*], C. Jin[1,2], B. Huang[1,3], N. K. Fontaine[1], R. Ryf[1], K. Shang[1], N. Grégoire[2], S. Morency[2], R.-J. Essiambre[1], G. Li[3,4], Y. Messaddeq[2] and S. LaRochelle[2]

1 Nokia Bell Labs, 791 Holmdel Road, Holmdel, New Jersey 07733, USA.
2 Center for Optics, Photonics and Lasers, Université Laval, Québec G1V0A6, Canada.
3 CREOL, The College of Optics & Photonics, University of Central Florida, Orlando, Florida 32816, USA.
4 The College of Precision Instruments and Opto-electronic Engineering, Tianjin University, Tianjin 300072, China.
*e-mail: haoshuo.chen@nokia-bell-labs.com

Abstract:

Space-division multiplexing (SDM), whereby multiple spatial channels in multimode[1] and multicore[2] optical fibres are used to increase the total transmission capacity per fibre, is being investigated to avert a data capacity crunch[3,4] and reduce the cost per transmitted bit. With the number of channels employed in SDM transmission experiments continuing to rise, there is a requirement for integrated SDM components that are scalable. Here, we demonstrate a cladding-pumped SDM erbium-doped fibre amplifier (EDFA) that consists of six uncoupled multimode erbium-doped cores. Each core supports three spatial modes, which enables the EDFA to amplify a total of 18 spatial channels (six cores × three modes) simultaneously with a single pump diode and a complexity similar to a single-mode EDFA. The amplifier delivers >20 dBm total output power per core and <7 dB noise figure over the C-band. This cladding-pumped EDFA enables combined space-division and wavelength-division multiplexed transmission over multiple multimode fibre spans.

The bandwidth demands for optical networks are growing exponentially and will soon exceed the maximum achievable capacity of a single-mode fibre (SMF) due to fibre nonlinearities[4], resulting in a 'capacity crunch'[3] in the near future. The simplest way to avert the 'capacity crunch' is to use multiple SMF transmission systems in parallel. However, this most basic form of SDM requires duplication of all the optical amplifiers, reconfigurable optical add–drop multiplexers and transponders for each additional fibre, and is subject to linear scaling of cost and complexity. Alternatively, SDM that uses the spatial domain for integration can provide a similar capacity increase but with great potential for reducing the cost per bit of transmission. Examples include fibres with enlarged cores that support multiple spatial modes[1,5], fibres with multiple cores[2,6], wavelength switches that switch all cores/modes simultaneously[7] and optical fibre amplifiers[8,9].

In the past five years, tremendous progress has been made in both SDM components and SDM transmission. Spatial degrees of freedom have increased from 3 modes to 15 in few-mode fibres (FMFs) or multimode fibres (MMFs), and from 3 cores to 36 cores in multicore fibres (MCFs). Over 2 Pbit s$^{-1}$ transmission over 100 spatial channels in multimode MCFs[10,11] have been demonstrated. Different types of spatial multiplexers (SMUXs) based on bulk optics[12], fused fibre bundles[13], laser-inscribed three-dimensional waveguide devices[14] and photonic integration[15] have enabled efficient connections between multiple SMFs and SDM fibres. Efforts to reduce receiver computational complexity due to multiple-input-multiple-output (MIMO) processing include FMFs with low differential-group-modal-delay and low inter-mode group coupling[16,17] as well as MCFs with <−40 dB core-to-core crosstalk[18,19].

As for SDM optical amplifiers, core-pumped multimode and multicore EDFAs have previously been used for wavelength-division multiplexing (WDM)/SDM[20,21]. However, core pumping requires almost the same number of pump diodes as the number of cores, and offers only a slight reduction in complexity and costs compared to duplicating SMF amplifiers. Cladding-pumped amplifiers can significantly reduce the complexity and cost of multi spatial channel amplifiers[2,22-26]. Additionally, the gain for different modes can be simply balanced by employing uniform cladding pumping[27] instead of tailoring the pumping spatial mode content using multiple pumps in different modes[28] or incorporating complex doping profiles into the fibre design[29]. In cladding-pumped amplifiers, the pump light is coupled to the cladding modes independently from the core modes, which eliminates wavelength combining elements and also enables simple pumping schemes. For instance, side pumping[30] does not require access to the fibre facet, but instead couples pump light into the cladding modes through the side of the fibre (Supplementary Section 'Side pumping'). The pump light illuminates all cores (Fig. 1a), eliminating the need for a separate pump diode per core. Moreover, uniform illumination of the cores simplifies modal gain equalization. The main drawback that has to be overcome with cladding-pumped amplifiers is the reduced pump power conversion efficiency, because the illuminated cladding area is much larger than the core area.

In this Letter, we explore integrated SDM optical amplifiers that can amplify many spatial channels and therefore significantly reduce the complexity and cost of SDM transmission systems. We have designed and fabricated an annular-cladding six-core erbium-doped fibre (EDF; Fig. 1b) to increase the pump power conversion efficiency. This work demonstrates amplification of 18 spatial channels (six cores with three spatial modes each) using only a single optical pump. The inner cladding is a depressed index region able to prevent pump light from entering the central region due to total internal reflection. This enhances the pump intensity around the cores in the annular cladding and can save more than 25% pump power compared to a uniform cladding (Supplementary Section 'Pump power for strong population inversion'), enabling a >20 dBm output power per core and <7 dB noise figure (NF) over the entire C-band (Supplementary Section 'Challenges in cladding-pumped EDFA'). This amplifier was then used in a 120 km WDM/SDM transmission experiment composed of three amplified spans.

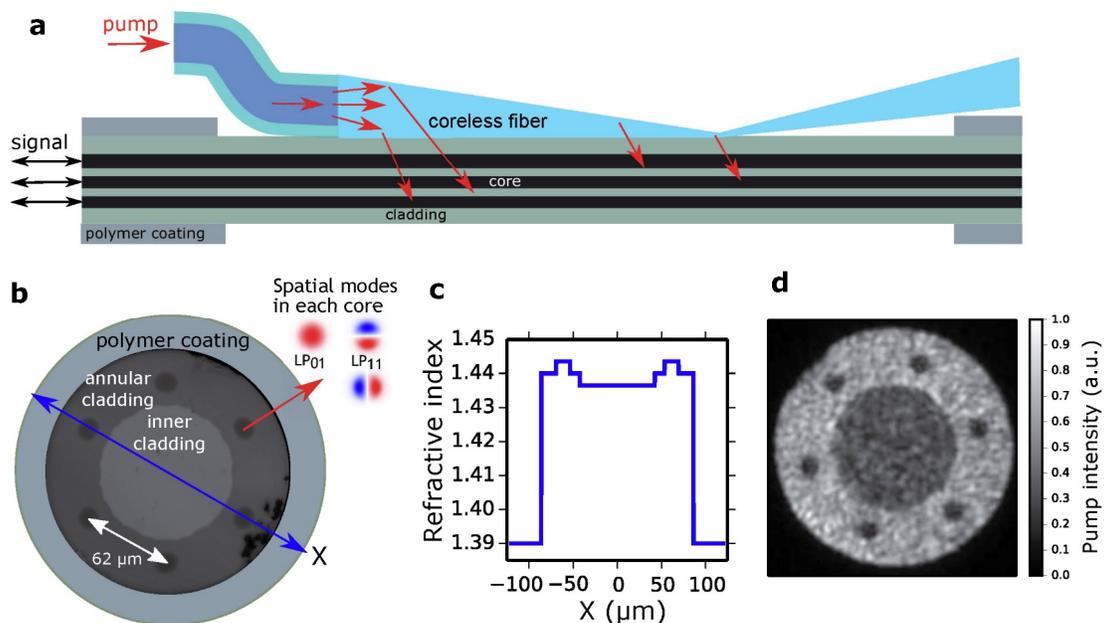

Figure 1 | Cladding-pumped six-core EDFA. a, Schematics of side pumping. b, Image of EDF facet. c, Refractive index profile of EDF. d, Output pump intensity distribution with multimode side pumping at 980 nm.

The refractive index profile along the x axis (Fig. 1b) of the annular-cladding six-core EDF is presented in Fig. 1c. The fabrication of the six-core EDF is addressed in detail in Supplementary Section 'Fibre preparation and fabrication'. Each core supports the $LP_{01}$ mode and the two degenerated $LP_{11}$ modes in the C-band (Supplementary Section 'Fibre param-eters'). To achieve high output power per spatial channel, which can only be accomplished if the EDFA maintains a high ratio of pump intensity to signal intensity throughout the gain medium so that it is not saturated, we designed core and cladding refractive indices to provide larger mode areas, thereby reducing the signal intensity without affecting the pump intensity. The calculated mode field areas are 168 and 179 $\mu m^2$ for the $LP_{01}$ and $LP_{11}$ mode groups, respectively, which are much larger than those of SMFs. The nearly identical mode fields also contribute to minimizing the mode-dependent gain, which was measured to be less than 2 dB (Supplementary Fig. 6c). We demonstrated that the average pump intensity in the annular cladding is enhanced by a factor of 1.45 compared to the inner cladding (Fig. 1d), which contributes to lower NFs through full population inversion at the input of the amplifier. It was verified that the annular cladding has negligible impact on the side-pump coupling efficiency, which is around 96%, similar to that of the uniform cladding (Supplementary Section 'Ray-tracing simulation').

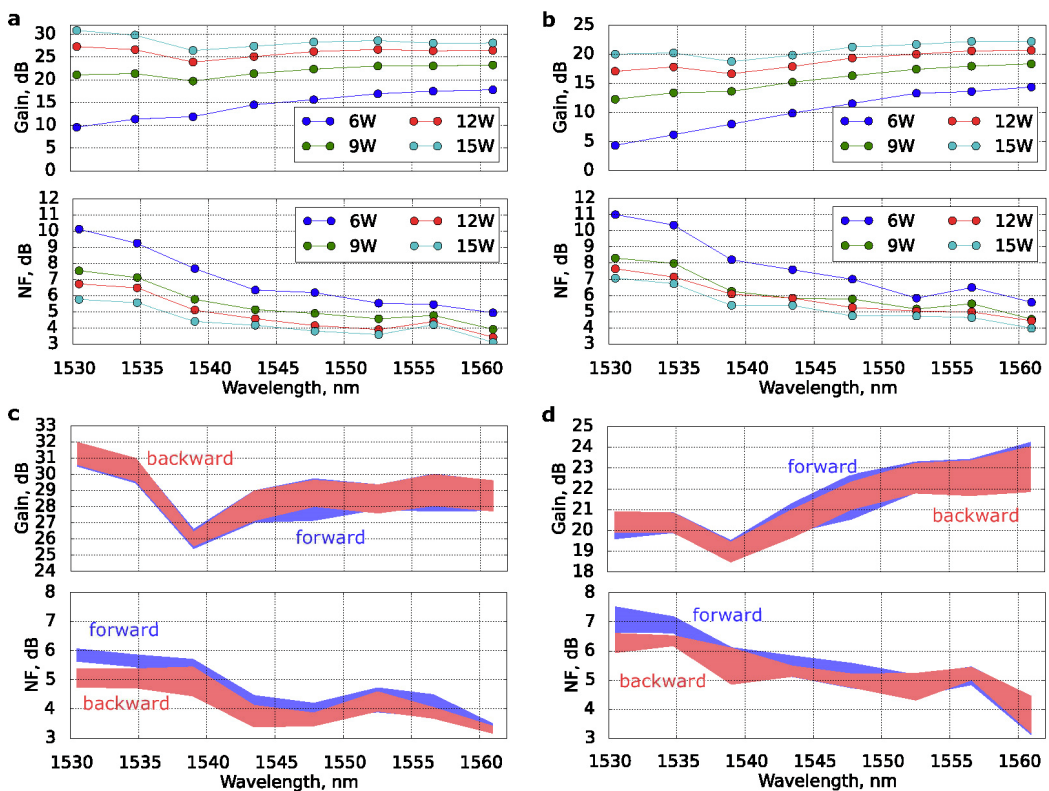

Figure 2 | Internal gain and NF characterization. a,b, Results for $LP_{01}$ mode of one core in a forward-pumping configuration with a total input power of −12 and −2 dBm, respectively, as a function of coupled-pump power. c,d, Range of internal gains and NFs in both forwards and backwards pumping for all six cores with a total input power of −12 and −2 dBm, respectively.

Figure 2a,b shows the internal gains (excluding input and output coupling losses) and NFs for one of the six cores as a function of coupled-pump power in the forward-pumping configuration (Supplementary Section 'Gain and noise figure measurements'). Figure 2c,d shows the range of internal gains and NFs for all six cores with a coupled-pump power of 15 W and total input signal power of −12 and −2 dBm, respectively. The cladding-pumped six-core EDFA provides an output power of >20 dBm per core and <7 dB NF over the C-band for the $LP_{01}$ mode. Lower NFs at shorter wavelengths can be achieved by further increasing the pump intensity to saturate the amplifier less. The gain and NF deviations of 1 dB are mainly attributed to the inaccuracy in the measurements of input and output coupling losses. Performance discrepancies, especially for NFs at short wavelengths, can be observed between forward and backward pumping, which can be attributed to inadequate pump power for forward pumping at the input where the coreless fibre starts to be wrapped around the EDF. This slight deficiency can be eliminated by either using bidirectional pumping or by first coupling the pump into a passive fibre with an identical structure to the EDF. We replaced the SMF at the EDF output with a 50/125 μm MMF to detect the output power for all three spatial modes, as all the modes are excited at the same time, each with −2 dBm input power. It was measured that 22 dBm output power can be provided by each core with 15 W pump power coupled into the cladding.

To use each core as an independent amplifier, it is essential to have negligible core-to-core crosstalk and pump-depletion-induced crosstalk. The worst-case core-to-core crosstalk for a six-core EDFA spliced with two tapered fibre bundles was measured to be better than −35 dB, which is mainly attributed to fabrication errors in the tapered fibre bundles (Supplementary Section 'Crosstalk') and can be further minimized through fabrication optimization. The ratio between the single-core area and cladding area is 1.2%. A benefit of this small overlap of the gain medium with the pump is negligible pump depletion, which avoids pump-depletion-induced crosstalk. No pump-depletion-induced crosstalk was observed under different signal loading conditions (Supplementary Fig. 6c).

Figure 3a presents an overview of the multi-span FMF transmission experiment using the cladding-pumped six-core EDFA. The 120 km fibre link contains four graded-index FMF spans that support six spatial and polarization modes at 1,550 nm. The effective areas were 64 and 67 μm$^2$ for the $LP_{01}$ and $LP_{11}$ modes, respectively. The absorption was 0.226 dB km$^{-1}$ and the chromatic dispersion was 18.5 ps nm$^{-1}$ km$^{-1}$ at 1,550 nm. The different group delay varied from 60 to 140 ps km$^{-1}$ for different spans. Figure 3b depicts how all six amplifying cores were used for either forward or backward transmission. In the forwards direction, a different core was used for amplification between the FMF spans (cores 1, 2 and 6). In the backwards direction, for the three remaining cores, dummy channels were launched to fully load the amplifier. The six FMFs of each fan-out were spliced to the FMF spans. The amplifier was operated with 8 W fibre-coupled power to compensate for the losses from the transmission FMF and the fan-outs.

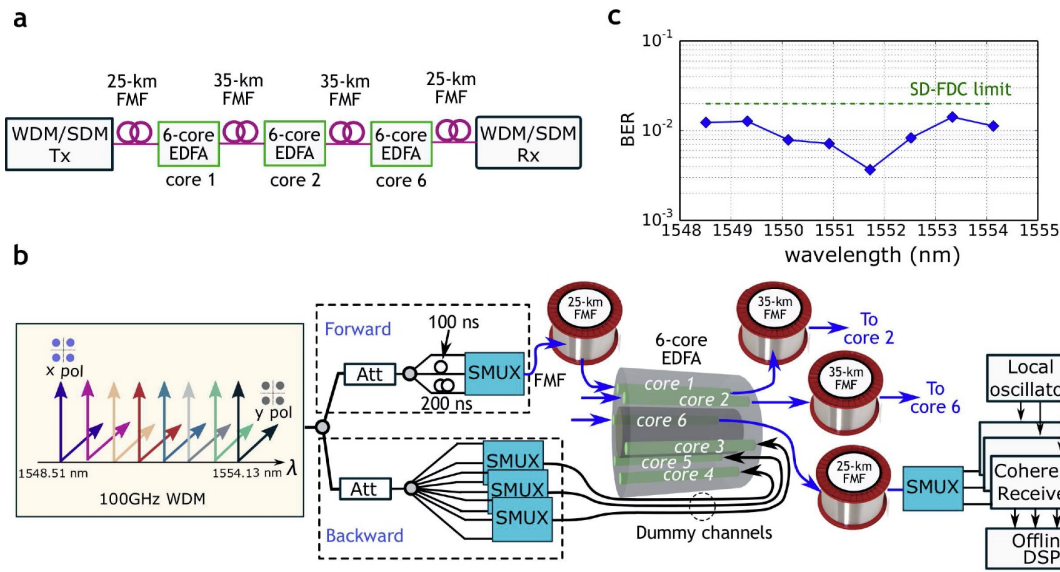

Figure 3 | Multi-FMF span transmission. a, Overview of multi-FMF span with four FMFs and three amplifier units. b, The transmission set-up includes a WDM/SDM transmitter for generating signal and dummy channels, a 120 km transmission link with one in-line six-core EDF and a SDM receiver. c, BER for eight WDM channels.

The transmitter, which produced eight wavelengths spaced at 100 GHz, is illustrated in Fig. 3b. The even and odd wavelengths were separately modulated by independent 30 Gbaud quadrature phase shift keyed (QPSK) waveforms with Nyquist pulse shaping, and the WDM comb was then polarization multiplexed. Twelve independent launch signals were prepared by splitting and delay decorrelating the WDM comb and were coupled into the six spatial and polarization modes of the FMFs using mode-group selective photonic-lantern-based SMUXs. Optical attenuators were used to equalize the input signal power into each core. The combined SDM and WDM forward signal was transmitted through three cores of the amplifier over 35 km FMF after the first and second stage, and 25 km FMF after the transmitter and another 25 km FMF in front of the receiver. At the receiver, the forward SDM signals were demultiplexed by another photonic-lanternbased SMUX and fed to three polarization-diversity-multiplexed coherent receivers operating at 40 GS

s⁻¹. MIMO-based offline digital signal processing (DSP) was applied to recover the signals (Supplementary Section 'Data processing').

Figure 3c shows the bit-error rate (BER) curves for all wavelength channels averaged across all spatial modes after the 120 km FMF span, which are below the $2 \times 10^{-2}$ soft-decision forward error correction (SD-FEC) limit for all wavelengths with a net capacity of 2.4 Tb s⁻¹ assuming 20% overhead. After 120 km transmission, the optical-signal-to-noise ratio (OSNR) was larger than 17 dB (compared to the OSNR of 35 dB at the transmitter output), which provides an equivalent NF of 6 dB per core considering fibre losses and coupling losses (Supplementary Section 'Equivalent noise figure'). The transmission reach was constrained by the accumulated mode-dependent loss (MDL) around 26 dB, induced at each splicing point between the six-core EDF and the fan-in/fan-out due to mode-profile mismatch and pitch deviation after down-tapering (Supplementary Section 'Mode-dependent loss'). By optimizing the tapering parameters and using fibres with identical mode profiles, we should be able to minimize the accumulated MDL in addition to employing differential group delay-compensated FMF spans to increase the transmission reach.

This work presents a novel type of EDF capable of amplifying 18 spatial channels using both cores and modes, which is scalable to support a larger spatial channel count. More than 20 dBm total output power per core and <7 dB NF was achieved through cladding pumping, which can be an efficient and low-cost solution for integrated optical amplification. Increasing the number of guided modes for each core is an efficient means by which to increase the spatial channel counts within a limited cladding diameter, and negligible mode-dependent gain can be achieved by oversizing the core to support more modes than required[27]. The depressed index region is not limited to the centre and can be separated and distributed to any location of the cladding to enhance pump intensity. By increasing the refractive index contrast between the undepressed cladding region containing cores and the depressed index region, or by using an air hole for the depressed index region, the pump power conversion efficiency can be further enhanced. Future research needs to determine the maximum number of cores, the modes per core, and address the optimum ratio between core area and undepressed cladding area to maximize the pump conversion efficiency without inducing core-to-core pump-depletion-induced crosstalk.

Acknowledgements

This work was supported in part by the ICT R&D program of MSIP/IITP, Republic of Korea (R0101-15-0071, 'Research of mode-division-multiplexing optical transmission technology over 10 km multimode fibre'), by the National Basic Research Program of China (973, project no. 2014CB340103/4), by the Canada Research Chair in Advanced Photonics Technologies for Communications (APTEC), by the Canada Excellence Research Chair in Enabling Photonic Innovations for Information and Communications (CERCP) and the Natural Sciences and Engineering Research Council of Canada (NSERC), and by NSFC Projects 61377076, 61307085 and 61335005. The authors acknowledge OFS Labs for the few-mode fibre. The authors also thank R.W. Tkach and P.J. Winzer for support and valuable discussions.

Author contributions

H.C. and C.J. developed the concept. C.J. and S.L. designed the fibre. N.G., S.M. and Y.M. fabricated the fibre. H.C., C.J., B.H. and K.S. conducted the fibre amplifier characterization. H.C. and C.J. conducted the ray tracing simulation. H.C., B.H. and N.K.F. fabricated the fan-in/fan-out. H.C., N.K.F. and R.R. conducted the transmission experiments. H.C. and N.K.F. wrote the manuscript. R.-J.E., G.L., Y.M. and S.L. helped write the article and provided funding.